\documentclass[preprint2]{aastex}

\newcommand{\myemail}{shogo@kusastro.kyoto-u.ac.jp}

\slugcomment{Version 080916}

\shorttitle{GC Polarization}
\shortauthors{Nishiyama et al.}

\begin{document}

\title{Magnetic Field Configuration at the Galactic Center
Investigated by Wide Field Near-Infrared Polarimetry}

\author{Shogo Nishiyama\altaffilmark{1,2},
Motohide Tamura\altaffilmark{2},
Hirofumi Hatano\altaffilmark{3},
Saori Kanai\altaffilmark{3},
Mikio Kurita\altaffilmark{3},
Shuji Sato\altaffilmark{3},
Noriyuki Matsunaga\altaffilmark{1},
Tetsuya Nagata\altaffilmark{1},
Takahiro Nagayama\altaffilmark{1},
Ryo Kandori\altaffilmark{2},
Yasushi Nakajima\altaffilmark{2},
Nobuhiko Kusakabe\altaffilmark{4},
Yaeko Sato\altaffilmark{4},
James H. Hough\altaffilmark{5},
Koji Sugitani\altaffilmark{6},
and Haruyuki Okuda\altaffilmark{7}
}

\altaffiltext{1}{Department of Astronomy, Kyoto University, 
Kyoto 606-8502, Japan; \myemail}

\altaffiltext{2}{National Astronomical Observatory of Japan, 
Mitaka, Tokyo 181-8588, Japan}

\altaffiltext{3}{Department of Astrophysics, Nagoya University, 
Nagoya 464-8602, Japan}

\altaffiltext{4}{Department of Astronomical Sciences,
Graduate University for Advanced Studies (Sokendai),
Mitaka, Tokyo 181-8588, Japan}

\altaffiltext{5}{Centre for Astrophysics Research, University of Hertfordshire,
Hatfield, Herts AL10 9AB, UK}

\altaffiltext{6}{Graduate School of Natural Sciences, Nagoya City University,
Nagoya 467-8501, Japan}

\altaffiltext{7}{Institute of Space and Astronautical Science, 
Japan Aerospace Exploration Agency, 
Sagamihara, Kanagawa 229-8510, Japan}

\begin{abstract}

We present a polarimetric map of a $20\arcmin \times 20\arcmin$ area
toward the Galactic center.
The polarization of point sources has been measured 
in the $J$, $H$, and $K_S$ bands
using the near-infrared polarimetric camera SIRPOL on the 1.4 m telescope IRSF.
One percent or better accuracy 
of polarization degree is achieved
for sources with $J<14.5$, $H<13.5$, and $K_S<12.0$.
Comparing the Stokes parameters between high extinction stars 
and relatively low extinction ones,
we have obtained a polarization originating from magnetically aligned dust grains
at the central region of our Galaxy of at most 1$-$2 kpc.
The distribution of 
the position angles shows a peak at $\sim$20\degr,
nearly parallel to the Galactic plane,
suggesting a toroidal magnetic configuration.
The derived direction of the magnetic field is in good agreement with
that obtained from far-infrared/submillimeter observations,
which detect polarized thermal emission from dust in the molecular clouds 
at the Galactic center.
Our results show that
by subtracting foreground components,
near-infrared polarimetry allows investigation of
the magnetic field structure {\it at } the Galactic center.

\end{abstract}



\section{INTRODUCTION}
\label{sec:Intro}

The discovery of polarized radio emission extending perpendicular
to the Galactic plane,
such as nonthermal radio filaments \citep{Yusef84} 
and polarized plumes \citep{Tsuboi86},
has provided early evidence for 
a substantial poloidal component of the magnetic field
in the central region of our Galaxy.
High resolution observations at radio wavelengths
have revealed more filaments of similar orientations
\citep[e.g.,][]{LaRosa04,Yusef04}, 
where the magnetic fields were found to be predominantly
aligned along the filaments.
The accumulation of these observational results has led to the hypothesis
that most of the volume of the Galactic center (GC) is permeated by a {\it poloidal} magnetic field.

Evidence for the existence of a {\it toroidal} magnetic structure
comes from far-infrared (FIR) and submillimeter (sub-mm) observations,
which detect polarized thermal emission from magnetically aligned dust grains.
The rotation axis of the dust grain aligns with the magnetic field, and thus
the polarization of the thermal emission indicates the magnetic field direction
(the measured direction of the E-vector is orthogonal to the magnetic field).
The magnetic fields inferred from these observations
\citep[e.g.,][]{Werner88,Morris92,Hildebrand93}
run parallel to the Galactic plane.
Recent linear polarization observations of the 450-$\mu$m continuum
suggest a large-scale toroidal magnetic field 
extending over a region of $170 \times 30 $ pc \citep{Novak03}.

\citet{Chuss03} found an interesting dependence of 
the magnetic field direction on the sub-mm flux.
In low density regions, the field aligns generally
perpendicular to the Galactic plane,
while in high density regions, the field has a toroidal geometry.
One explanation for this is that
the global magnetic field in the early Galaxy
was initially in a poloidal configuration,
however the gravitational energy density in dense molecular clouds
was strong enough to distort the poloidal field
into a toroidal one,
though it was insufficient in the low density regions.

A model which can connect the poloidal and toroidal magnetic fields
was proposed by \citet{Uchida85},
and was extended to a more realistic case by \citet{Shibata86}.
This magnetohydrodynamic model was developed to explain 
the GC lobes found by \citet{Sofue84}.
Since magnetic flux is generally frozen into matter,
differential rotation and infall can shear an initially poloidal field 
into a toroidal one.
Consequently, a toroidal field is developed close to the Galactic plane,
while the field is vertical at high Galactic latitude.

The \citet{Uchida85} model predicts that
the toroidal component will generally be more dominant
closer to the Galactic plane.
A map of the direction of the magnetic field
could therefore be a simple indicator of 
how well this model works in the GC.
However, dust emissivity is high in dense, warm clouds, and thus
observations of dust emission in FIR/sub-mm wavelengths
are strongly limited to such regions,
which show patchy distributions in the GC region.

In this paper, we present near-infrared (NIR) polarimetry
of point sources toward the GC
covering a much larger region of the sky than 
previous observations of this region.
We demonstrate that NIR polarization
can provide information on the magnetic field structure
not only in the Galactic disk, 
but also in the central region of our Galaxy.

\section{OBSERVATIONS AND DATA REDUCTION}
\label{sec:Obs}

We conducted NIR polarimetric observations of the GC
with the SIRPOL camera on the night of 4 July 2006.
SIRPOL consists of a single-beam polarimeter 
\citep[a half-wave plate rotator unit and a fixed wire-grid polarizer;][]{Kandori06}
and the NIR imaging camera SIRIUS
\citep[Simultaneous Infrared Imager for Unbiased Survey;][]{Nagas99, Nagay03},
and is attached to the 1.4-m telescope IRSF (Infrared Survey Facility).
SIRPOL provides images of a 7\farcm7 $\times$ 7\farcm7 area of sky 
in three NIR wavebands, $J$ ($1.25\mu$m), $H$ ($1.63\mu$m),
and $K_S$ ($2.14\mu$m), simultaneously.
The detectors are three 1024 $\times$ 1024 HgCdTe arrays,
with a scale of 0\farcs45 pixel$^{-1}$.  
The filter system of IRSF/SIRPOL is similar to 
the MKO system \citep{Tok02}.

We observed a $20\arcmin \times  20\arcmin$ area (nine SIRPOL fields)
centered at the position of Sgr A$^*$
($17^{\mathrm{h}} 45^{\mathrm{m}}  40.0^{\mathrm{s}}$,
$-29\degr 00\arcmin 28\farcs0 $; J2000.0).
We took 10-s exposures each at 4 wave plate angles 
($0\fdg0$, $22\fdg5$, $45\fdg0$, $67\fdg5$)
at each of 10 dithered positions.
The weather condition was photometric, with a seeing of 
$\sim$1\farcs2 ($J$),
$\sim$1\farcs1 ($H$),
and $\sim$1\farcs0 ($K_S$).
The IRAF (Image Reduction and Analysis Facility)\footnote{
IRAF is distributed by the National Optical Astronomy
Observatory, which is operated by the Association of Universities for
Research in Astronomy, Inc., under cooperative agreement with
the National Science Foundation.}
software package was used
to perform dark- and flat-field corrections,
followed by sky background estimation and subtraction.

\section{DATA ANALYSIS AND RESULTS}

\subsection{Polarization of Point Sources}
\label{sec:PolPoint}

To obtain the photometric magnitudes and errors in the three bands,
we used the DAOFIND task in the DAOPHOT package \citep{Stetson87} 
to identify point sources
in Stokes $I$ images [$I=(I_{0\degr}+I_{22\fdg5}+I_{45\degr}+I_{67\fdg5})/2$].
The sources were then input 
to the ALLSTAR task for PSF-fitting photometry.
About 10 sources were used to construct the PSF in each image.
Each Stokes $I$ image was calibrated with 
the photometric image of the same position obtained in previous imaging observations
\citep{Nishi06a},
in which the standard star \#9172 \citep{Persson98} was used for calibration.
We assumed that \#9172 has magnitudes of
$J=12.48$, $H=12.12$, and $K_S=12.03$ in the IRSF/SIRIUS system.

\begin{figure}[h]
 \begin{center}
   \plotone{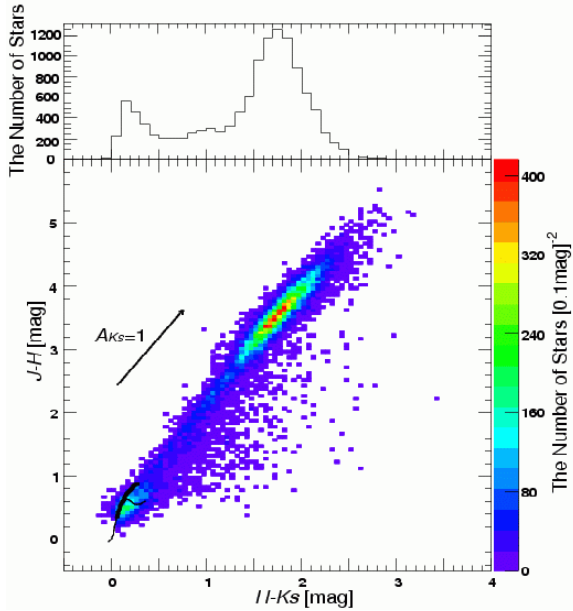}
    \caption{
      $H-K_S$ histogram ({\it top}) and 
      $JHK_S$ color-color diagram ({\it bottom}) for point sources 
      with $J$, $H$, and $K_S$ photometric errors of less than 0.11 mag.
      The thick and thin curves represent the loci
      of giants and dwarfs, respectively \citep{Tokunaga00}.
      The arrow indicates the $A_{K_S}=1$ mag reddening vector \citep{Nishi06a}.
    }
  \label{fig:Col2}
 \end{center}
\end{figure}

In Fig. \ref{fig:Col2} 
we show the $H-K_S$ histogram (top panel) and 
the $JHK_S$ color-color diagram (bottom panel)
for stars with photometric errors of less than 0.11 mag.
Also plotted are loci of unreddened giants and dwarfs.
The arrow is parallel to the reddening vector,
and its length corresponds to $A_{K_S}=1$ mag
\citep{Nishi06a}.
Considering large extinction toward the GC,
stars with small $H-K_S$ should be attributed to foreground (disk) stars,
and stars consisting of the strongest peak in the $H-K_S$ histogram
is attributed to ones in the Galactic bulge.

Astrometric calibration was performed, field by field,
with reference to the positions of point sources
in the 2MASS point source catalog \citep{Skrutskie06}.
Sources with photometric errors of less than 0.05 mag
in 2MASS and our catalog were used for the calibration.
As a result of this astrometric calibration,
we obtained an rms of the positional difference of better than 0\farcs1
for sources with a $< 0.11$ mag photometric error.

The Stokes parameters $I$, $Q$, and $U$ for point sources
were determined from aperture polarimetry of combined images as follows. 
DAOFIND and APPHOT tasks were used for the point-source identification 
and the aperture photometry.
We then obtained the intensity for each wave plate angle
($I_{0\degr}$, $I_{22\fdg5}$, $I_{45\degr}$, $I_{67\fdg5}$).
Since aperture photometry with a small aperture radius
gives a better photometric result than PSF-fitting photometry,
aperture photometry was applied in the following procedure.
The size of the PSF is slightly different among the four images due to variations of the seeing,
hence we used different apertures for each image.
The aperture diameters were set to be equal to
$2 \times \mathrm{FWHMs}$ of the best fit Gaussian profile (GFWHM)
determined by the PSFMEASURE task.
The means of the adopted aperture sizes were 
$\sim$1\farcs2 ($J$),
$\sim$1\farcs1 ($H$),
and $\sim$1\farcs0 ($K_S$).
The position angle offset of SIRPOL $\alpha$ was estimated to be  $\alpha = 105\degr$ \citep{Kandori06}
which was set to the origin of the angles $0\degr$, $22\fdg5$, $45\degr$, and $67\fdg5$.
Based on the intensities of the four angles,
we calculated the total intensity $I$ and two ``raw'' Stokes parameters $Q'$ and $U'$ as
\[I = (I_{0\degr} + I_{22\fdg5} + I_{45\degr} + I_{67\fdg5})/2,\]
\[Q' = I_{0\degr} - I_{45\degr},\]
\[U' = I_{22\fdg5} - I_{67\fdg5}.\]
The polarization degree $P$ and the position angle $\theta$ were derived by 
\[ P = \sqrt{(Q'^2+U'^2)}/I, \]
\[ \theta = \frac{1}{2} \arctan(U/Q), \]
where
\begin{eqnarray}
Q = Q' \cos(2 \alpha) - U' \sin(2 \alpha), U = Q' \sin(2 \alpha) + U' \cos(2 \alpha).
\label{eq:AngCor}
\end{eqnarray}
The debiased $P$ was finally derived by $P_{\mathrm{db}} = \sqrt{P^2-\delta P^2}$
where $\delta P$ is the error of $P$, given by
\begin{eqnarray}
 \delta P = \frac{1}{P} \sqrt{ \left( \frac{Q'}{I} \right)^2 \left[ \delta \left( \frac{Q'}{I} \right) \right]^2 +
\left( \frac{U'}{I} \right)^2 \left[ \delta \left( \frac{U'}{I} \right) \right]^2 }. 
\label{eq:PError}
\end{eqnarray}
We regard sources for which $(P^2-\delta P^2) \leq 0$
as non-polarized source,
and we do not consider such sources further in this paper.
The typical magnitudes for $\delta P = 1 \%$ are
14.5 ($J$), 13.5 ($H$), and 12.0 ($K_S$).

\begin{figure}[]
 \begin{center}
   \epsscale{0.9}
   \plotone{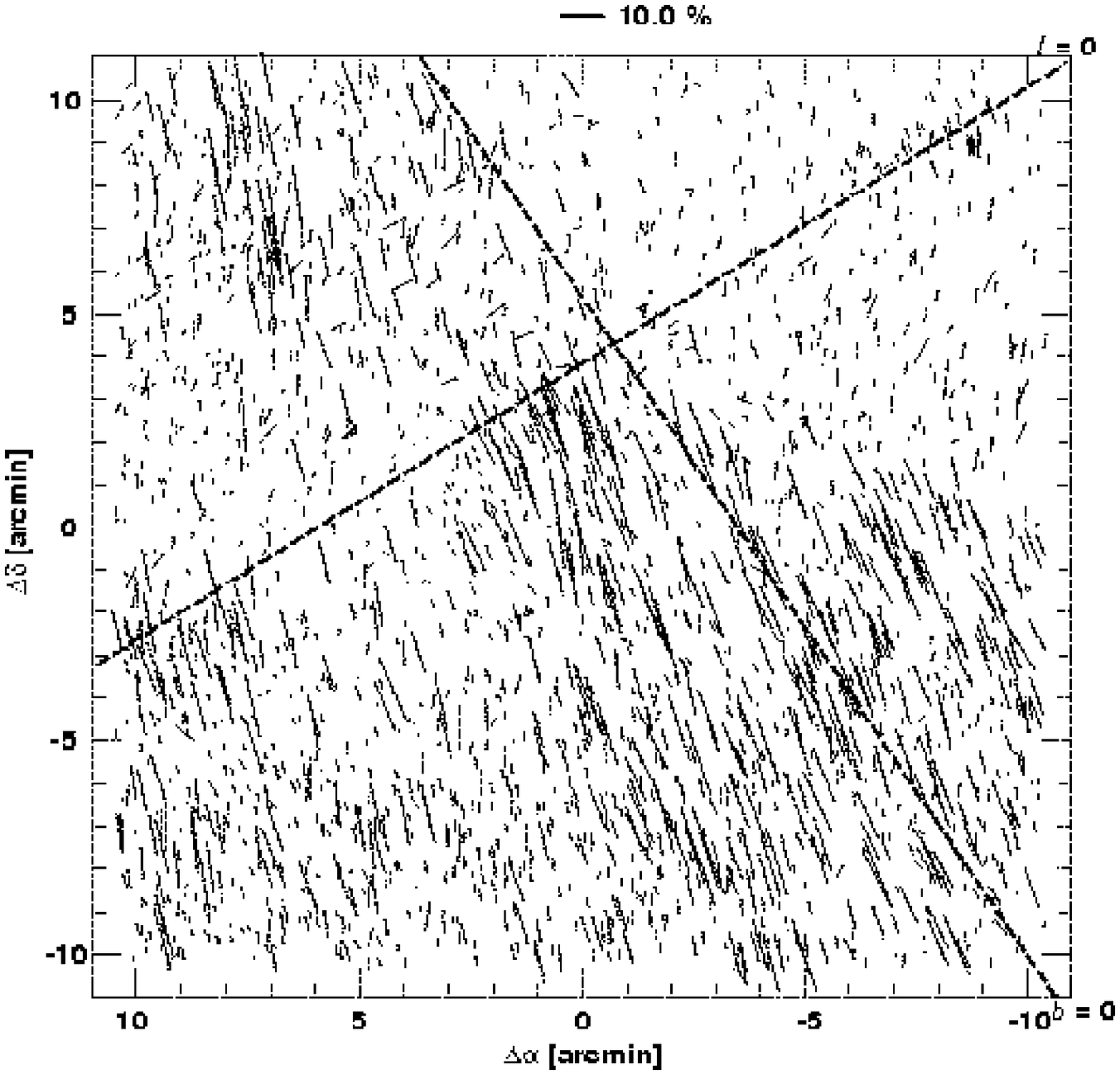}
   \plotone{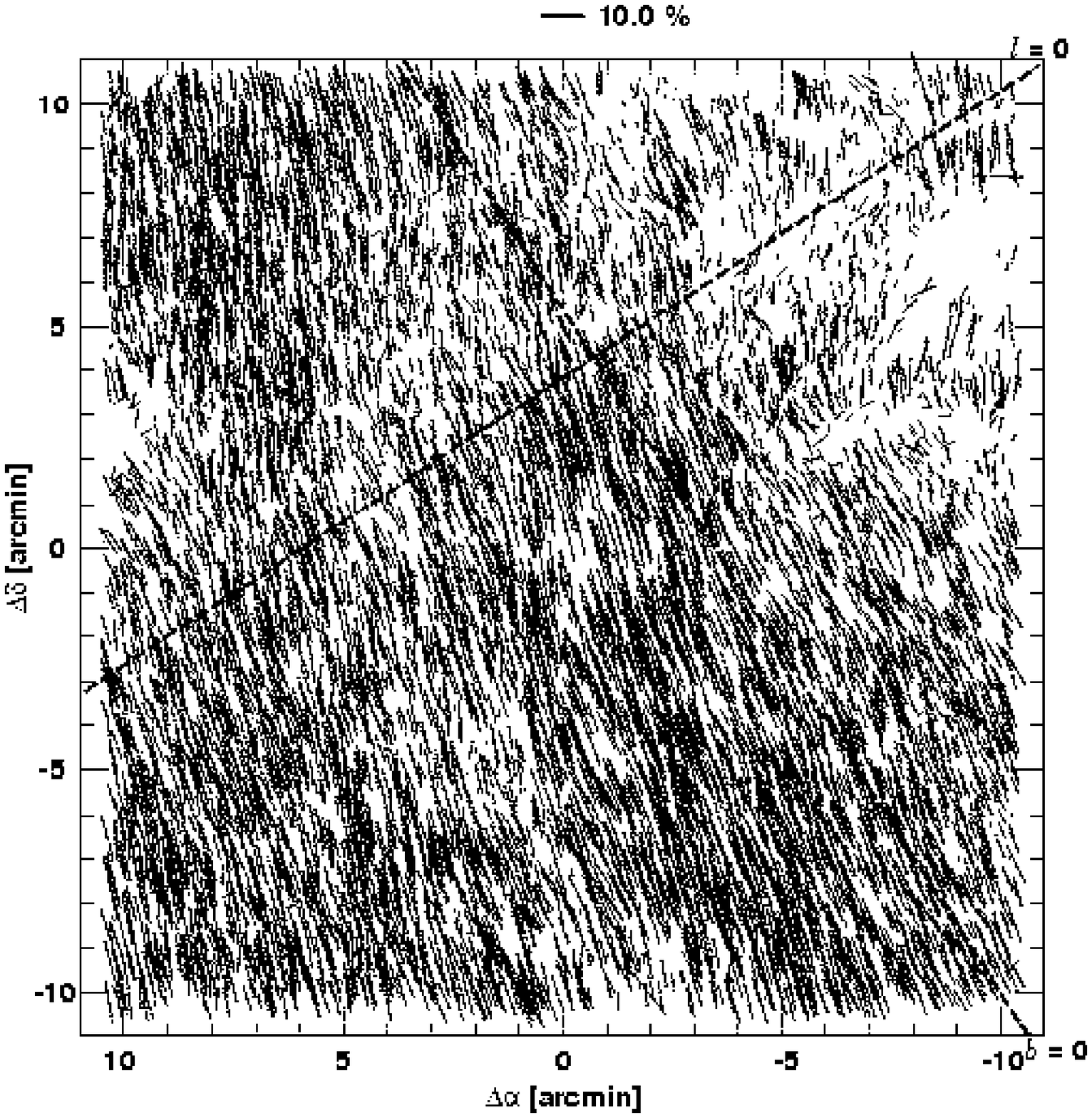}
   \plotone{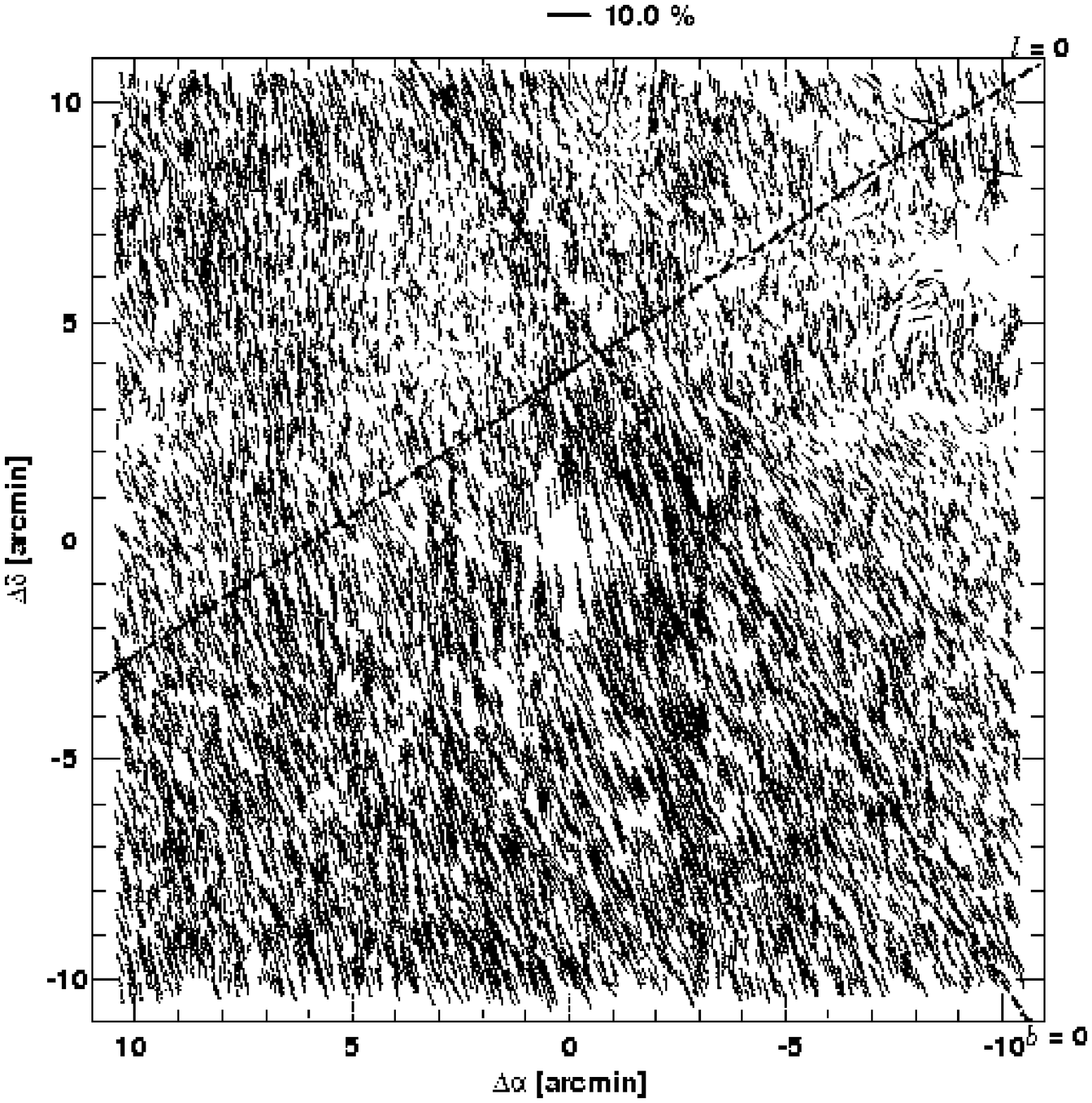}
    \caption{
      Polarization of the Galactic center 
      for stars with $P > 0\%$ and $\delta P < 1\%$.
      2243, 7963, and 9661 stars 
      in the $J$ (top), $H$ (middle), and $K_S$ (bottom) bands, respectively,
      are plotted.
      The coordinate offsets (\arcmin) were measured with respect to
      the location of Sgr A$^*$.
      Each bar is drawn parallel to the E-vector of the measured polarization.
      Their length indicates the measured degree of polarization.
    }
  \label{fig:Vmap}
 \end{center}
\end{figure}

Fig. \ref{fig:Vmap} plots 
the degree and position angle of stars
with a degree of polarization determined to have an accuracy better than 1\%.
The orientation of each bar gives the inferred direction of polarization
and the length of the bar is proportional to the degree of polarization.
The coordinate offsets (\arcmin) were measured with respect to the location of Sgr A$^*$.

\begin{figure}[h]
 \begin{center}
  \rotatebox{180}{
    \plotone{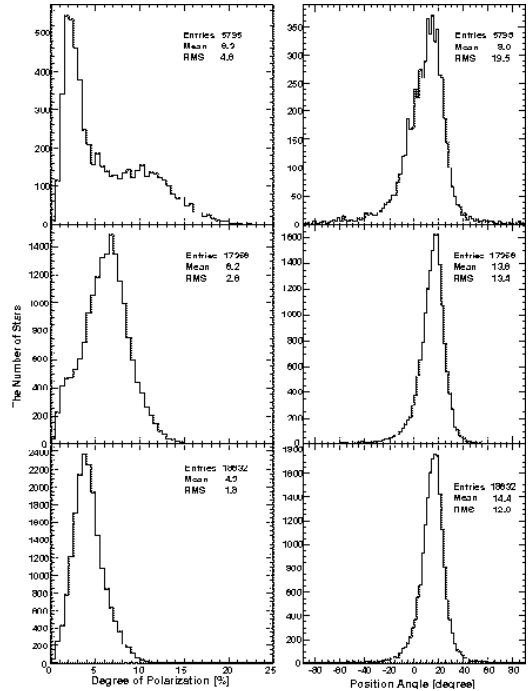}
  }
    \caption{
      Histograms of degree of polarization (left) and position angle (right)
      in the $J$ (top), $H$ (middle), and $K_S$ (bottom) bands,
      for stars with $P > 0\%$ and $\delta P < 3\%$. 
     5795 ($J$), 17356 ($H$), and 18632 ($K_S$) stars are employed in these histograms.
    }
  \label{fig:HistAP}
 \end{center}
\end{figure}

Histograms of debiased $P$ and $\theta$ in each band 
are shown in Fig. \ref{fig:HistAP}.
The mean degree and angle are
6.3 \% and 8\fdg0 in the $J$ band,
6.2 \% and 13\fdg6 in the $H$ band,
and 4.3 \% and 14\fdg4 in the $K_S$ band.
\citet{Kob80} detected $K$-band emission of unresolved point sources
at the central $7\arcmin \times 7\arcmin$ region
and obtained an average degree of polarization of 5\%,
with  position angles of 10\degr ~to 15\degr,
showing a good agreement with our results in the $K_S$ band.

\begin{figure}[h]
 \begin{center}
  \rotatebox{90}{
    \plotone{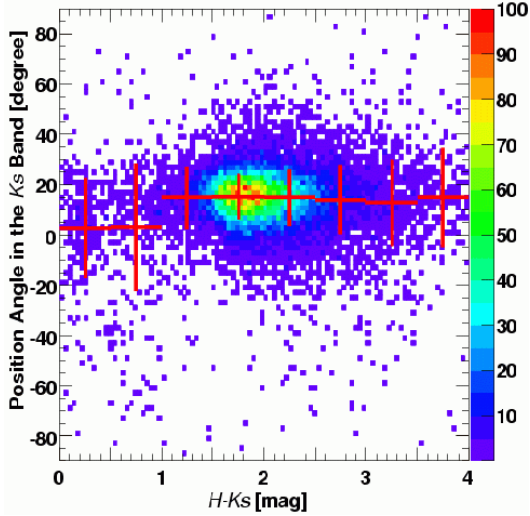}
  }
    \caption{
      Position angle $\theta$ in the $K_S$ band vs. $H-K_S$ diagram
      for stars with 
	  $\delta H < 0.11$, $\delta K_S < 0.11, P_{K_S} > 0\%$ and $\delta P_{K_S} < 3\%$. 
      The red crosses represent the mean and standard deviations of $\theta$
      in 0.5 mag width bins.
    }
  \label{fig:DistHKPK}
 \end{center}
\end{figure}

The correlation between $\theta$ in the $K_S$ band
and $H - K_S$ color is shown in Fig. \ref{fig:DistHKPK}.
To clarify the dependence of $\theta$ on $H - K_S$,
we divided the $H - K_S$ data set in bins of equal size (0.5 mag)
and calculated the mean and the standard deviation of $\theta$ in each bin,
represented by red crosses in Fig. \ref{fig:DistHKPK}.
We can see a change of the mean position angle at $H - K_S \sim 1.0$:
the mean angle is $\sim 5\degr$ at $0 < H - K_S < 1.0$,
while it is $\sim 15\degr$ at $H - K_S > 1.0$.
Such a change has already been indicated by \citet{Kob83},
and is also in good agreement with their results.

We have identified two distinct populations in Fig. \ref{fig:HistAP}:
stars with small $P$ and small $\theta$ 
(typically $P_J \la 5\%$ and $\theta \la 0\degr$), 
and stars with larger $P$ and $\theta \ga 10\degr$
(see also Fig. \ref{fig:DistHKPK}).
It is conceivable that the former are nearby stars,
and the latter are stars distributed in the Galactic bulge.
The stars that corresponds to 
the strong peak in the $P_{J}$ histogram at $\sim 2\%$
have a color of $H-K_S \sim$ 0.1$-$0.2 (see also Fig. \ref{fig:Col2}) and $\theta_{J} \sim 0\degr$ 
and thus are most likely to be nearby dwarfs.
This peak can be found in the $H$ band around $P_H \sim 1 \%$,
but becomes invisible in the $P_{K_S}$ histogram.
In the $K_S$ band, most of the stars detected
are red giants in the Galactic bulge,
which constitute a strong peak
in the $P_{K_S}$ histogram.
The strong peak at $P_{K_S} \sim 4\%$
corresponds to those at $\sim 7\%$ in $P_{H}$,
and at $\sim 10\%$ in $P_{J}$,
because this change of degree of polarization can be explained by
the power law $P_{\lambda} \propto \lambda^{-2}$ 
of the interstellar polarization \citep{Nagata94}.
The distinct populations,
and the wavelength dependence of the polarization will be discussed in another paper
(H. Hatano et al., in preparation).

\begin{figure}[h]
 \begin{center}
  \rotatebox{90}{
  \plotone{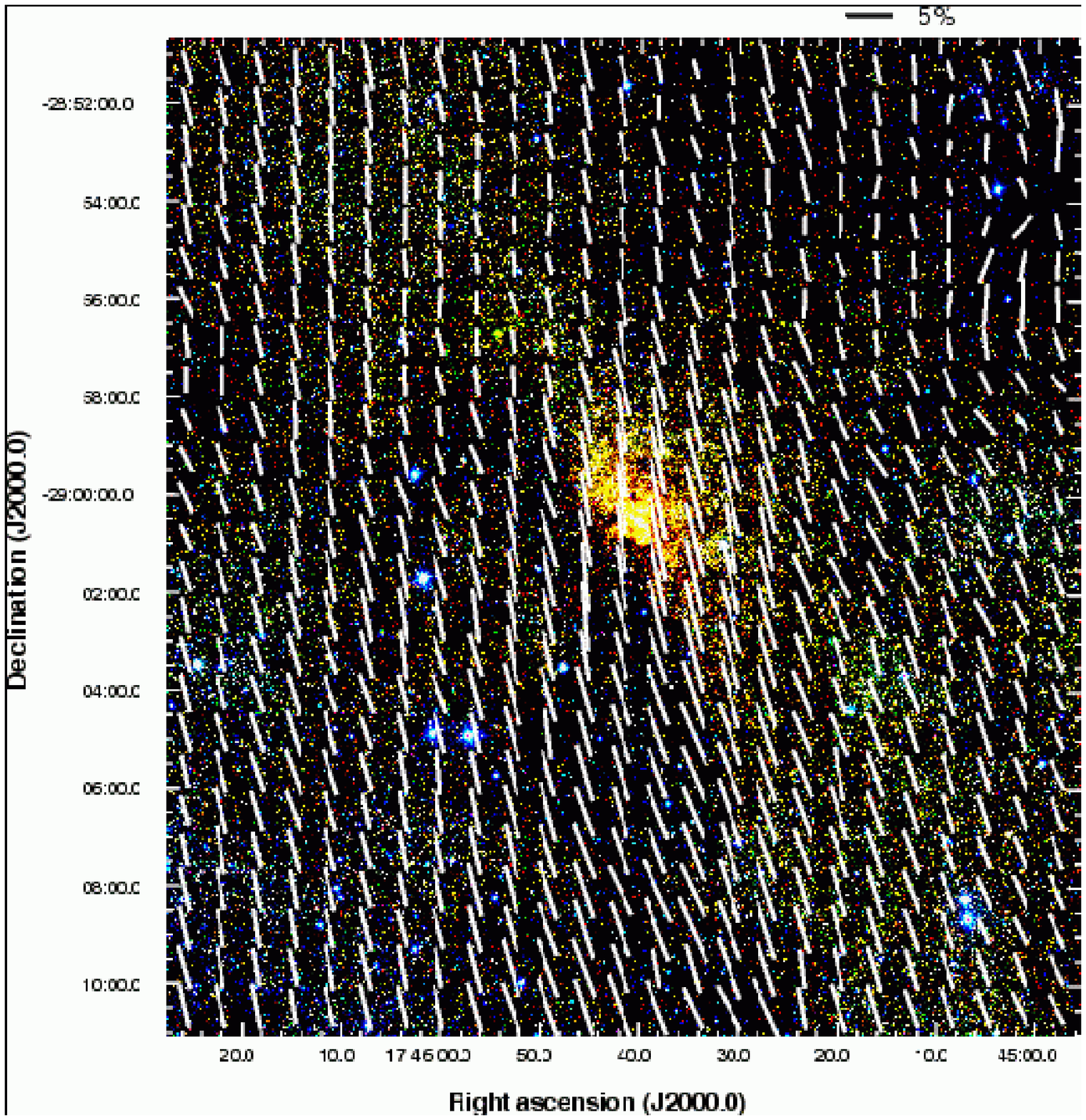}
  }
    \caption{
      $K_S$-band polarization vector map superposed on 
      the three color ($J$, $H$, $K_S$) composite image of the Galactic center.
      The Galactic center is the bright yellow blob in the center.
      The mean $P_{K_S}$ and ${\theta}_{K_S}$ are calculated 
      for each $0\farcm8 \times 0\farcm8$ grid.
      Note here that stars with
      $P_{K_S} > 0\%$ and ${P_{K_S}}/\delta P_{K_S}>3.0$
      are used for the calculation, so that
      the mean $P_{K_S}$ might be overestimated.
    }
  \label{fig:VHmapCol}
 \end{center}
\end{figure}

Since the vector maps in Fig. \ref{fig:Vmap} 
are crowded and thus almost illegible,
we show the $K_S$-band mean vector map in Fig. \ref{fig:VHmapCol}.
The mean degree and position angles were calculated 
using stars in a $0\farcm8 \times 0\farcm8$ grid
with ${P_{K_S}}/\delta P_{K_S}>3.0$.
The vectors are superposed on 
the three color ($J$, $H$, $K_S$) composite image of the same region.
At a first glance, 
most of the vectors are in order,
which can also be seen in the $\theta_{K_S}$ histogram,
and are nearly parallel to the Galactic plane.
Moving north-eastward across the image,
the position angles slightly rotate clockwise.
At a few positions where the number density of stars is small,
and hence strong foreground extinction exists,
the vectors have irregular directions,
particularly at the northwestern corner.
These irregularities might be explained by
the inherent magnetic field configuration in foreground dark clouds.

\subsection{Separating Foreground Polarization and Galactic Center Component}

Polarization vectors of stars trace the plane-of-the-sky projection of the magnetic field,
and polarimetric measurements of stars of different distances reveal
the three-dimensional distribution of the magnetic field orientations.
In our observed fields, we can detect stars in the Galactic disk and bulge.
From the stars at the close side in the Galactic bulge 
(referred to hereafter as ``blue stars'' due to their relatively small reddening),
we can obtain the degree of polarization and position angle,
which are affected mainly by interstellar dust in the Galactic disk.
The light from stars at the far side in the bulge
(hereafter ``red stars'')
is transmitted through the dust in the disk {\it and} the bulge.
Therefore, using both blue and red stars,
we can obtain the bulge (GC) component of the polarization.
The procedure is as follows.
(In the following process, only $K_S$-band polarization of stars
is used.)

\begin{figure}[h]
 \begin{center}
    \rotatebox{-90}{
      \epsscale{1.0}
      \plotone{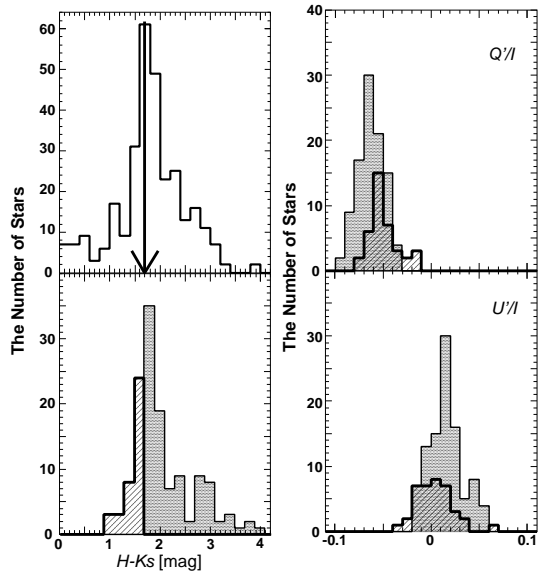}
    }
    \caption{
      Upper left: $H-K_S$ histogram of a sub-field ($l,b = +1\arcmin, -1\arcmin$)
      with stars with $\delta H < 0.11$, $\delta K_S < 0.11$, and $H \leq 15.0$.
      The arrow represents the peak of the histogram.
      Lower left: $H-K_S$ histograms of the same sub-field
      for blue (hatched) and red (dotted) stars
      with $P_{K_S} > 0\%$ and $\delta P_{K_S} < 3\%$. 
      Upper and lower right: $Q'/I$ and $U'/I$ histograms for  
      the blue (hatched) and red (dotted) stars
      shown in the lower left histogram.
    }
  \label{fig:ProcGCMag}
 \end{center}
\end{figure}

\begin{figure}[h]
  \begin{center}
      \plotone{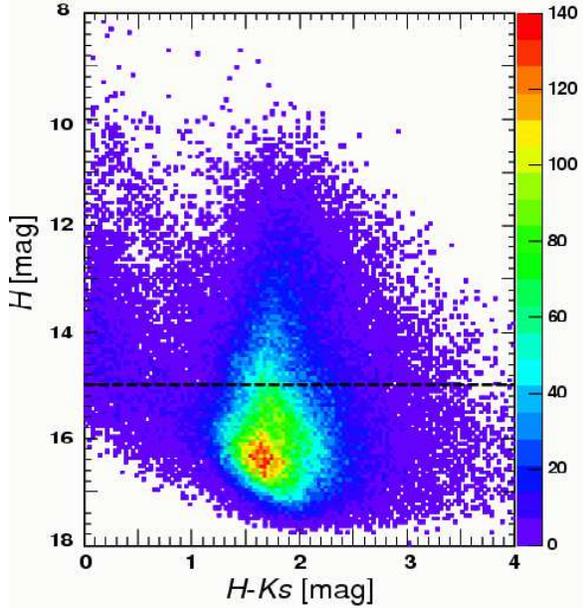}
    \caption{
      $H$ vs. $H - K_S$ color-magnitude diagram
      for stars with $\delta H < 0.11$ and $\delta K_S < 0.11$.
      The dashed line represents $H = 15.0$.
    }
    \label{fig:CMDAll}
  \end{center}
\end{figure}

As a first step,
we divided the field into $10 \times 10$ sub-fields 
of $2\arcmin \times 2\arcmin$ and 
drew $H-K_S$ histograms for each sub-field with stars of 
$\delta H < 0.11$, $\delta K_S < 0.11$, and $H \leq 15.0$,
which are much brighter than the 10$\sigma$ limiting magnitude
in the $H$ band (16.7 mag).
One of the $H-K_S$ histograms is shown in Fig. \ref{fig:ProcGCMag},
upper left panel.
Using the histograms, we evaluated
a peak value of the histogram ${(H-K_S)}_{\mathrm{peak}}$
for each sub-field.
Fig. \ref{fig:CMDAll} is an $H$ vs. $H-K_S$ color-magnitude diagram (CMD)
for stars with $\delta H < 0.11$ and $\delta K_S < 0.11$.
This CMD shows that the criterion $H \leq 15.0$ is bright enough
to avoid the influence of the limiting magnitudes ($H \approx 16.7$, $K_S \approx 15.5$)
on the determination of ${(H-K_S)}_{\mathrm{peak}}$.

\begin{figure}[h]
  \begin{center}
      \plotone{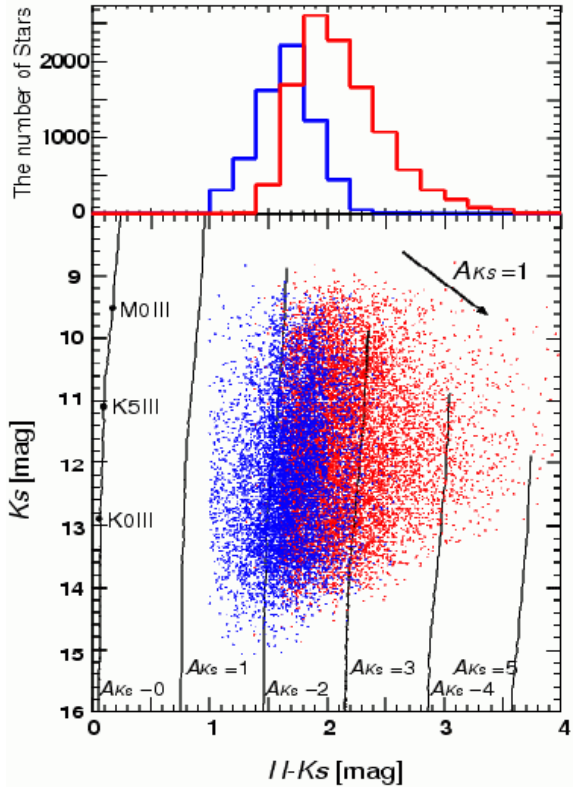}
    \caption{
      $H - K_S$ histograms (top) 
      and $K_S$ vs. $H - K_S$ color-magnitude diagram (bottom)
      for red and blue stars with $P_{K_S} > 0\%$ and $\delta P_{K_S} < 3\%$.
      Theoretical isochrones for different extinctions
      of $A_{K_S} = 0, 1, 2, 3, 4,$ and 5 mag,
      are shown by solid curves.
      The arrow indicates the $A_{K_S}=1$ mag reddening vector \citep{Nishi06a}.
    }
    \label{fig:CMDRedBlue}
  \end{center}
\end{figure}

Using the $H-K_S$ color, we divided the stars 
with $\delta P < 3\%$ into three sub-groups:
``nearby'' stars and ``blue'' and ``red'' stars in the bulge.
We assume that nearby stars have a color of $H-K_S<1.0$,
because at $H-K_S \sim 1.0$
the number of stars drops and approaches a minimum (Fig. \ref{fig:Col2}, top panel)
and a clear change of position angles can be determined (Fig. \ref{fig:DistHKPK}).
The ``blue'' stars are redder than $H-K_S=1.0$ and 
bluer than ${(H-K_S)}_{\mathrm{peak}}$.
The stars with $H-K_S > {(H-K_S)}_{\mathrm{peak}}$
are selected as ``red'' stars
(lower left panel in Fig. \ref{fig:ProcGCMag}).
The average and standard deviation of ${(H-K_S)}_{\mathrm{peak}}$ 
of the 100 sub-fields are 1.85 and 0.24, respectively;
Fig. \ref{fig:CMDRedBlue} shows 
the $H - K_S$ histogram (top panel) of the red and blue stars,
and their location in the $K_S$ vs. $H - K_S$ CMD (bottom panel).
The blue stars have a peak at $H - K_S \approx 1.5$,
while the red stars have a peak at $H-K_S \approx 2.0$.
The red and blue stars of {\it all} sub-fields are plotted, 
so their distribution is overlapped in the CMD of Fig. \ref{fig:CMDRedBlue}.
There are $\sim 60$ to $\sim 340$ stars with $\delta P$ better than 3\%
in each sub-field in the $K_S$ band.

As a second step to subtract the polarization originating in the bulge, 
$Q'/I$ and $U'/I$ histograms in the $K_S$ band were constructed
for the blue and red stars in each sub-field
(upper and lower right panels in Fig. \ref{fig:ProcGCMag}).
We calculated their means as
$<Q'/I>_{\mathrm B}$, $<Q'/I>_{\mathrm R}$,
$<U'/I>_{\mathrm B}$, and $<U'/I>_{\mathrm R}$.
We then obtained the degree of polarization and position angle for the blue stars,
\[
P_{\mathrm B} = \sqrt{ \left< \frac{Q'}{I} \right>^{2}_{\mathrm B} 
  + \left< \frac{U'}{I} \right>^{2}_{\mathrm B} },~~
\theta_{\mathrm B} = \frac{1}{2} \arctan \left( 
\frac{\left< \frac{U}{I} \right>_{\mathrm B}}{\left< \frac{Q}{I} \right>_{\mathrm B}} \right),
\]
and for the red stars,
\[
P_{\mathrm R} = \sqrt{ \left< \frac{Q'}{I} \right>^{2}_{\mathrm R} 
  + \left< \frac{U'}{I} \right>^{2}_{\mathrm R} },~~
\theta_{\mathrm R} = \frac{1}{2} \arctan \left( 
\frac{\left< \frac{U}{I} \right>_{\mathrm R}}{\left< \frac{Q}{I} \right>_{\mathrm R}} \right).
\]
where 
$< U/I >_{\mathrm B}$ and $< Q/I >_{\mathrm B}$, and 
$< U/I >_{\mathrm R}$ and $< Q/I >_{\mathrm R}$
are the Stokes parameters 
whose position-angle offset are corrected [see equations (\ref{eq:AngCor})].
We obtained $P$ and $\theta$ for the red minus blue components
using the following equations \citep{Goodrich86}:
\[
P_{\mathrm {R-B}} = \sqrt{ \left( \left< \frac{Q'}{I} \right>_{\mathrm R} - \left< \frac{Q'}{I} \right>_{\mathrm B} \right)^{2} 
  + \left( \left< \frac{U'}{I} \right>_{\mathrm R} - \left< \frac{U'}{I} \right>_{\mathrm B}  \right)^{2} }, 
\]
\[
\theta_{\mathrm {R-B}} = \frac{1}{2} \arctan \left( 
\frac{\left< \frac{U}{I} \right>_{\mathrm R} - \left< \frac{U}{I} \right>_{\mathrm B}}
     {\left< \frac{Q}{I} \right>_{\mathrm R} - \left< \frac{Q}{I} \right>_{\mathrm B}} \right),
\]
where 
$(< U/I >_{\mathrm R} - < U/I >_{\mathrm B})$
and 
$(< Q/I >_{\mathrm R} - < Q/I >_{\mathrm B})$
are also the Stokes parameters 
whose position-angle offset are corrected [see equations (\ref{eq:AngCor})].
The errors of $<Q'/I>$ and $<U'/I>$ were calculated from the standard error on the mean
$\sigma/\sqrt{N}$ of the $Q'/I$ and $U'/I$ histograms,
where $\sigma$ is the standard deviation 
and $N$ is the number of stars.
The error of $P_{\mathrm {R-B}}$ was then calculated by propagation of errors 
[see equation (\ref{eq:PError})].

\begin{figure}[h]
 \begin{center}
    \plotone{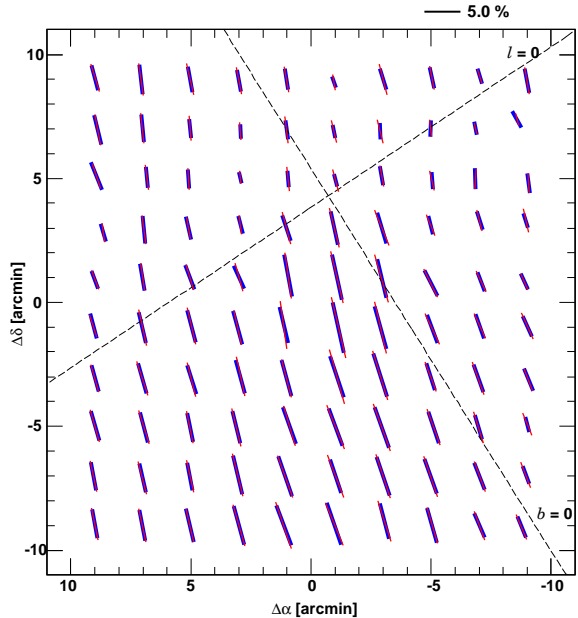}
    \caption{
      $K_S$-band polarization map derived from the blue-star component 
      ($P_{\mathrm B}$ \& $\theta_{\mathrm B}$ : {\it blue bars}) 
      and red-star component
      ($P_{\mathrm {R}}$ \& $\theta_{\mathrm {R}}$ : {\it red bars}).
    }
  \label{fig:VmapFGRed}
 \end{center}
\end{figure}

\begin{figure}[h]
 \begin{center}
  \epsscale{0.9}
    \plotone{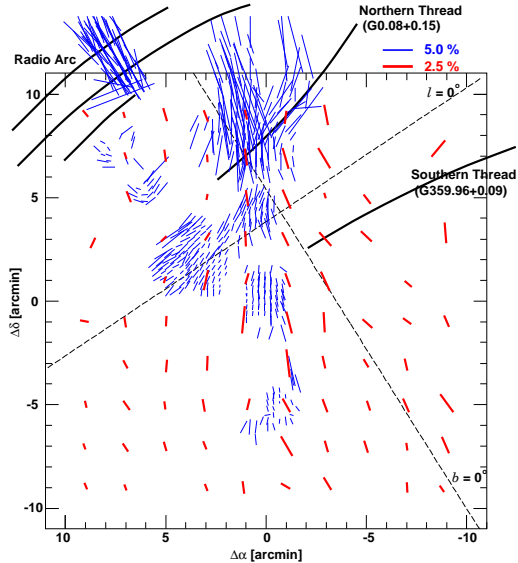}
    \caption{
      $K_S$-band polarization map derived from the Galactic center component
      ($P_{\mathrm {R-B}}$ \& $\theta_{\mathrm {R-B}}$, {\it red bars}),
      where the polarization is detected with $P_{\mathrm {R-B}}/\delta P_{\mathrm {R-B}} \geq 2$.
      The polarization map at the center of our Galaxy derived 
      from FIR/sub-mm observations ({\it blue bars}) is also shown.
      The length of the bars is proportional to the measured
      degree of polarization,
      and their orientation is drawn parallel to the inferred
      magnetic field direction.
      The FIR/sub-mm wavelengths data sets are from 
      60 $\mu$m \& 100 $\mu$m polarimetry by \citet{Dotson00},
      and 350 $\mu$m polarimetry by \citet{Novak00} and 
      \citet[][see also their Fig. 1]{Chuss03}.
      Some prominent radio filaments are shown as heavy dark lines.
    }
  \label{fig:PABGChuss}
 \end{center}
\end{figure}

\begin{figure}[h]
 \begin{center}
  \plotone{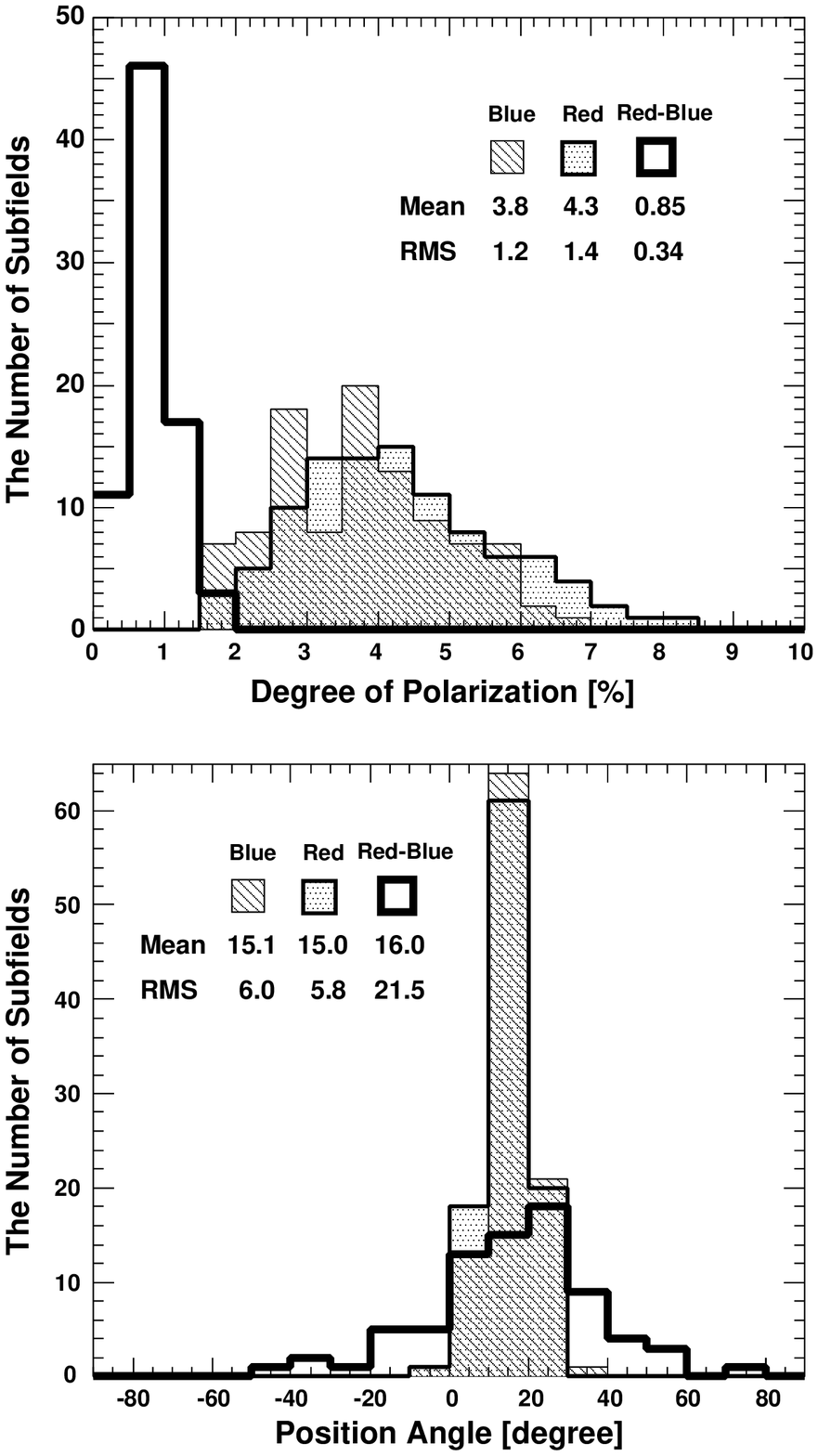}
    \caption{
      Top: Histograms of degrees of polarization 
      for $P_{\mathrm B}$ ({\it hatched}),
      $P_{\mathrm R}$ ({\it dotted}),
      and $P_{\mathrm {R-B}}$ ({\it white}).
      Bottom: Histograms of position angles
      for $\theta_{\mathrm B}$ ({\it hatched}),
      $\theta_{\mathrm R}$ ({\it dotted}),
      and $\theta_{\mathrm {R-B}}$ ({\it white}).
    }
  \label{fig:HistPAFGBG}
 \end{center}
\end{figure}

We show vector maps for $P_{\mathrm B}$ and $\theta_{\mathrm B}$ (blue bars) and
$P_{\mathrm R}$ and $\theta_{\mathrm R}$ (red bars) in Fig. \ref{fig:VmapFGRed},
and the same map for $P_{\mathrm {R-B}}$ and $\theta_{\mathrm {R-B}}$
in Fig. \ref{fig:PABGChuss}.
The average of $P_{\mathrm B}$ and $\theta_{\mathrm B}$ are
3.8 \% and 15\fdg1, 
and of $P_{\mathrm R}$ and $\theta_{\mathrm R}$ are
4.3 \% and 15\fdg0, respectively 
(Fig. \ref{fig:HistPAFGBG}, hatched and dotted histograms).
Those of $P_{\mathrm {R-B}}$ and $\theta_{\mathrm {R-B}}$ are
also obtained as 0.85 \% and 16\fdg0
(white histogram in Fig. \ref{fig:HistPAFGBG}),
only for grids where the polarization
is detected with $P_{\mathrm {R-B}}/\delta P_{\mathrm {R-B}} \geq 2$.
The averages for $\theta_{\mathrm B}$ and $\theta_{\mathrm {R-B}}$
are similar, but their dispersions are different:
the standard deviation of $\theta_{\mathrm B}$ is 6\fdg0, 
while that of $\theta_{\mathrm {R-B}}$ is 21\fdg5
(the average of $\delta \theta_{\mathrm {R-B}}$ is 7\fdg6.)
The small dispersion of $\theta_{\mathrm B}$ suggests that
the long axis of the interstellar dust grains in the Galactic disk is well aligned
perpendicular to the Galactic plane.
The histogram of $\theta_{\mathrm {R-B}}$ has a peak at $\sim 20\degr$,
which also roughly coincides with the angle of the Galactic plane.
A similar result is also obtained for the $H$-band polarization.

In Fig. \ref{fig:PABGChuss},
our $P_{\mathrm {R-B}}$ and $\theta_{\mathrm {R-B}}$ are
plotted overlaid on the $B$-vectors derived from FIR/sub-mm observations
\citep{Dotson00,Novak00,Chuss03}.
Although the observations are restricted to the position of dense molecular clouds,
we find good agreement between the FIR/sub-mm and NIR observations
in spite of the difference of wavelengths and methods of deriving polarization.

\section{DISCUSSION}
\label{sec:Disc}

\subsection{Previous Infrared Polarimetry toward the GC}
\label{subsec:PrevNIR}

NIR polarimetry for diffuse emission and point sources toward the GC
has been conducted since the 1970's \citep{Maihara73}.
From the polarization angle aligned nearly along the Galactic plane,
and the wavelength dependence of polarization 
well fitted by a power law \citep{Nagata94},
it has been interpreted that the polarization is of interstellar origin
dominated by dust that lies in the Galactic disk
\citep[e.g.,][]{Hough78,Kob80,Lebofsky82}.

One of the deepest NIR polarimetric observations toward the GC was carried out 
by \citet{Eckart95}
for a small field of $\sim 13\arcsec \times 13\arcsec$.
They established significant polarization 
for 160 sources fainter than 13 mag in the $K$ band.
The mean flux-weighted polarization is 4 \% at 25\degr,
nearly parallel to the Galactic plane.
A similar result, $4.1 \pm 0.6 \%$ at $30 \pm 10\degr$, was also obtained by \citet{Ott99}. 
\citet{Eckart95} concluded that most of the polarizations are
caused via absorption by aligned dust grains in the Galactic plane.

A change of the magnetic field configuration 
along the line of sight toward the GC
has been pointed out by \citet{Kob83}.
The diagram of position angles in the $K$ band versus $H-K$ for 15 discrete sources
\citep[Fig. 5,][]{Kob83} showed that
the position angles are smaller and less ordered
for sources with $H-K<1.0$,
while those with $H-K>1.0$ are confined to a relatively narrow range around $20\degr$.
They concluded that this may indicate a change
of the magnetic field direction at a distance corresponding to $H-K=1.0$
(5 kpc or more according to their calculation).

We can also identify a change of the position angle 
in the $K_S$ bands at $H-K_S \sim 1.0$
(see Fig. \ref{fig:DistHKPK}).
As shown in the color-color diagram in Fig. \ref{fig:Col2},
most of the stars with $H-K_S \ga 1.0$ have the color of giants
(i.e., their unreddened positions are on the locus of giants),
and the strong peak in the $H-K_S$ histogram (the top panel in Fig. \ref{fig:Col2})
suggests that they are distributed in the Galactic bulge.
Hence, this change of the position angle indicates
a transition of the magnetic field configuration along the line of sight,
between the Galactic disk and bulge.

\subsection{``Red'' and ``Blue'' Stars and Their Separation in the Line of Sight}

To discriminate between foreground disk stars and those in the Galactic bulge,
the criterion $H-K_S=1.0$ is applied in our analysis
because most stars with $H-K_S>1.0$ are attributed as
giants in the Galactic bulge.
The histogram of $H-K_S$ and 
the color-color diagram of point sources 
with photometric errors of less than 0.1 mag in the three bands
are shown in Fig. \ref{fig:Col2}.
Nearby stars can be found around the giants' and dwarfs' loci
(thick and thin curves) without reddening,
while most of the redder stars, 
which are located at $H-K_S \ga 1.0$ and $J-H \ga 2.5$,
have a color expected for reddened giants.
Toward the GC, the number of stars in the bulge is 
larger than in the exponential disk by a factor of $\sim 50$ \citep{Wainscoat92},
and thus most of the stars
at $H-K_S \ga 1.0$ and $J-H \ga 2.5$
are giants in the Galactic bulge.

The bottom panel of Fig. \ref{fig:CMDRedBlue} shows 
the location of the red and blue stars
in the $K_S$ vs. $H - K_S$ color-magnitude diagram.
The theoretical Padova isochrones \citep{Girar02} for solar metallicity
with an age of 10 Gyr are also plotted in the color-magnitude diagram.
The isochrones are put at the distance of the GC \citep[7.5 kpc;][]{Nishi06b}
and shifted along the reddening vector \citep[$A_{K_S}/E_{H-K_S} = 1.44$;][]{Nishi06a}
with $K_S$ extinctions of 0, 1, 2, 3, 4, and 5 mag.
The distances to the red and blue stars are difficult to estimate accurately,
since we do not know the distribution of the interstellar dust
along the line of sight;
however, most of the giants in the Galactic bulge
can be detected for at least $K_S < 15$,
and for these stars,
we can find a clear peak in the $H-K_S$ histograms.
Fig. \ref{fig:CMDAll} clearly shows that
our observations are essentially equally sensitive
to stars on the near ($H-K_S \la 2.0$) and far ($H-K_S \ga 2.0$) sides
of the Galactic bulge for $H \leq 15.0$.
The clear peaks found in the $H-K_S$ histograms
are thus due to the spatial distribution of stars,
not due to a distance effect. 
This indicates that ${(H-K_S)}_{\mathrm{peak}}$ roughly
corresponds to a real peak of the spatial distribution of giants along the line of sight.
The color-magnitude diagram of Fig. \ref{fig:CMDRedBlue}  also tells us that
a large number of blue stars are distributed around the distance
where $1.0 \la A_{K_S} \la 2.0$,
and most of the red stars are further than the distance
which suffers from the extinction of $A_{K_S} = 2.0$.

From the distribution of $H-K_S$ color,
we estimate the depth of the region where
we have mapped the magnetic configuration shown in Fig. \ref{fig:PABGChuss}.
For each sub-field,
we calculated $H-K_S$ color differences 
between the peak in the $H-K_S$ histogram and the mean colors of blue and red stars,
i.e.,
$(H-K_S)_{\mathrm{peak}} - \langle (H-K_S)_{\mathrm{blue}} \rangle$ and
$\langle (H-K_S)_{\mathrm{red}} \rangle -(H-K_S)_{\mathrm{peak}}$.
These color differences and corresponding extinction 
$A_{{K_S}_{\mathrm{blue}}} = 1.44 \times [(H-K_S)_{\mathrm{peak}} - \langle (H-K_S)_{\mathrm{blue}} \rangle]$
and 
$A_{{K_S}_{\mathrm{red}}} = 1.44 \times [\langle (H-K_S)_{\mathrm{red}} \rangle - (H-K_S)_{\mathrm{peak}}]$,
where 1.44 comes from $A_{K_S}/E_{H-K_S} = 1.44$ \citep{Nishi06a},
show the amount of dust extinction in the region
where we have obtained polarimetric information.
First, the amount of extinction toward the peaks in the $H-K_S$ histograms
can be calculated to be 
$A_{{K_S}_{\mathrm{peak}}} = 1.44 \times [(H-K_S)_{\mathrm{peak}}-(H-K_S)_0]$,
where the mean intrinsic color $(H-K_S)_0$ of red giants is assumed to be $\sim 0.2$
(see Fig. \ref{fig:CMDRedBlue}).
This corresponds to the amount of extinction up to the GC in each sub-field.
Next, the extinctions $A_{{K_S}_{\mathrm{blue}}}$ and $A_{{K_S}_{\mathrm{red}}}$
are converted to the actual distances from the GC, using a simple model.
To make conservative estimates, we use the model by \citet{Davies97},
who derived a rather extended distribution of dust.
\citet{Davies97} showed that 
the extinction to the GC
associated with cool diffuse interstellar dust
can be calculated using 
\begin{eqnarray}
A = C \times \int_0^{R_0} e^{-r/\alpha_{\mathrm{d}}} dr,
\label{eq:Extinction}
\end{eqnarray}
where $\alpha_{\mathrm{d}} \approx 5.3$ kpc is
a scale-length of the dust distribution in the radial direction,
$C$ is a constant depending on the wavelength and dust density,
and $R_0$ is the distance between the GC and the Sun
\citep[$R_0 = 7.5$ kpc;][]{Nishi06b}.
The distances from the GC $x_{\mathrm{blue}}$ and $x_{\mathrm{red}}$
corresponding  to $A_{{K_S}_{\mathrm{blue}}}$ and $A_{{K_S}_{\mathrm{red}}}$, respectively,
can be estimated with the equations
\[ \Bigl( \int_0^{x_{\mathrm{blue,red}}} e^{-r/\alpha_{\mathrm{d}}} dr \Bigr) 
\Bigg/ \Bigl( \int_0^{R_0} e^{-r/\alpha_{\mathrm{d}}} dr \Bigr)
= A_{{K_S}_{\mathrm{blue,red}}} \Big/ A_{{K_S}_{\mathrm{peak}}}.\]
We obtained the average and standard deviation of $x_{\mathrm{blue}}$ as 0.5 and 0.1 kpc,
and those of $x_{\mathrm{red}}$ as 1.0 and 0.5 kpc, respectively.
We found a long-side tail in the $x_{\mathrm{red}}$ histogram,
which enlarges the average and the standard deviation.
This estimation suggests that
the polarization shown in Fig. \ref{fig:PABGChuss} occurs
between the average distances of $(R_0 - 0.5)$ kpc and $(R_0 + 1.0)$ kpc from the Sun,
arising probably from the central 1$-$2 kpc region of our Galaxy.

In reality, the central part of the Galaxy harbors
a strong concentration of gas and dust called
the ``Central Molecular Zone'' \citep[$R \la 200$ pc;][]{Morris96},
which approximately corresponds to the concentration of stars called
the ``Nuclear Bulge'' \citep[$R \la 300$ pc;][]{Mezger96,Serabyn96}.
According to \citet{Launhardt02}, stars belonging to the Nuclear Bulge
dominate in the central part of the Galaxy (see their Fig. 14).
Therefore, a large portion of the stars we have detected is
located in the Nuclear Bulge, and the polarization at the GC
originates mostly within the central few hundred pc.

\subsection{Magnetic Field Configuration at the GC}

As shown in Fig. \ref{fig:PABGChuss},
we present the magnetic field configuration at the central region of our Galaxy
by discriminating between the polarizations due to the disk and GC origin.
The peak of the histogram of the position angle is $\sim 20\degr$
(see Fig. \ref{fig:HistPAFGBG}),
almost parallel to the Galactic plane.
This coincidence, and 
the deficiency of $\theta_{\mathrm {R-B}}$ around $-60\degr$,
the angle perpendicular to the Galactic plane,
show the basically toroidal geometry of the magnetic field.
We cannot find a clear systematic dependence of position angle
on Galactic latitude (from $b \approx -0\fdg27$ to $+0\fdg18$),
suggesting that there is no systematic transition of 
the magnetic field direction in this region.

The direction of dust grain alignment at the GC has been investigated 
from polarized dust emission in the mid- and far-infrared and sub-mm wavelengths.
The magnetic field implied by the emission 
is generally parallel to the Galactic plane in the circumnuclear disk
\citep[e.g.,][]{Werner88,Morris92,Hildebrand93}.
As seen in Fig. \ref{fig:PABGChuss},
the magnetic field configuration we obtained at the GC 
shows a good agreement globally with those obtained by
\citet{Dotson00}, \citet{Novak00}, and \citet{Chuss03},
which are the highest angular resolution polarimetry data sets in the FIR/sub-mm wavelengths.
The local features also show an excellent agreement:
an X-shaped feature extends from $(\Delta \alpha, \Delta \delta) = (-3\arcmin, +10\arcmin)$
down through ($+5\arcmin, 0\arcmin$) 
\citep[described in \S 3.1.4,][]{Chuss03}
is also confirmed in our map.
The configuration at M$-$0.13$-$0.08 around ($-1\arcmin, -5\arcmin$) is also reproduced.
The polarized FIR/sub-mm emission comes from molecular clouds,
which are known to be located in the GC.
Therefore we conclude that the position angles derived from our NIR polarimetry
represent the direction of the aligned dust grains {\it in} the GC.

\citet{Chuss03} has suggested that the magnetic field aligns generally 
perpendicular to the Galactic plane in low density regions, 
while the field has a toroidal configuration in high density regions.
They explain this correlation using the idea that
in underdense regions, 
the magnetic field energy density can support itself against gravitational forces,
preserving a primordial poloidal magnetic field.
In overdense regions like molecular clouds, on the other hand, 
the gravitational forces are strong enough to shear the magnetic field
into a direction parallel to the Galactic plane.
In this context, lower density regions should have a poloidal configuration.
However, in our vector map,
the field shows a predominantly toroidal direction
even at locations where the FIR/sub-mm emission is too weak for polarimetry
[at the southeastern corner in Fig. \ref{fig:PABGChuss},
see also Fig. 1 in \citet{Chuss03} and Fig. 2 in \citet{Novak03}.]
The weak FIR/sub-mm emission suggests 
a paucity of dense clouds along the lines of sight,
and the polarization 
in this direction can be considered to be interstellar in origin.
Hence the ``interstellar'' magnetic field at the GC
is dominated by a toroidal configuration,
and the primordial poloidal magnetic field, if it existed,
is not preserved today in this region.

Several radio filaments exist in our observed field,
and three of them are prominent:
the GC Radio Arc \citep{Yusef84}, 
and the Northern and Southern Threads  
\citep[also known as G0.08+0.15 and G359.96+0.09;][]{Morris85}, 
which are shown in Fig. \ref{fig:PABGChuss}.
Polarization studies have confirmed that
the emission from the filaments is strongly linearly polarized,
and that the internal magnetic field orientations are
parallel to the long axes of the filaments \citep{Tsuboi86,Lang99}.
The simplest interpretation of these observations,
combined with the discovery of more filaments \citep[e.g.,][]{LaRosa04,Yusef04},
is that poloidal magnetic fields are
pervasive throughout the central few hundred pc.
Although the polarizations originating from the filaments cannot be detected in our observation,
those toward the surrounding fields can be detected.
The filaments have a width of less than $\sim 10\arcsec$ \citep{Lang99},
and probably have a depth similar to their width.
On the other hand, the size of the grids of our analysis is $2\arcmin \times 2\arcmin$,
and the polarization shown in Fig. \ref{fig:PABGChuss} is 
the average of $\sim$1$-$2 kpc from the GC along the line of sight.
Hence most of the stars we detected do not show a polarization originating from the filaments.
However, we can detect the average polarization near the line of sight toward the filaments.
As shown in Fig. \ref{fig:PABGChuss},
most of the position angles of the grids including the filaments 
align nearly perpendicular to them
rather than parallel.
Therefore, Fig. \ref{fig:PABGChuss} suggests that
the average magnetic field has a toroidal configuration 
even around the sight-lines toward the filaments.
We note again that the polarization is the average along the line of sight
and does not originate from the area close to the filaments.

\subsection{NIR Polarization as a New Tool for Mapping the GC Magnetic Field}

We have shown that the polarization of starlight can be 
a probe of the magnetic field near the GC.
\citet{Morris98} enumerated five different ways
in which the magnetic field near the GC has been studied:
morphology, polarization angle, Faraday rotation of the radio continuum,
Zeeman effect, and polarized dust emission in FIR/sub-mm wavelengths.
The polarization of starlight has not been employed
for such investigations.

The optical polarization of starlight can trace
the structure of the local magnetic field,
but cannot detect stars near the GC due to large extinction.
NIR polarimetry has been carried out toward the GC prior to our observations
(see \S \ref{subsec:PrevNIR}),
but division of polarization into several components
(e.g., originating from the Galactic disk and the central region)
has not been done previously.
The wide field-of-view of the NIR polarimeter SIRPOL,
and the statistical treatment of tens of thousands of stars
enable us to study the magnetic field near the GC;
that is, the NIR polarimetry of starlight is a new way 
to investigate the magnetic field.

NIR polarimetry has the advantage of
providing information about the magnetic field
at locations where FIR/sub-mm emission is weak.
NIR polarization of starlight is attributed 
to extinction along the line of sight by aligned dust grains,
while FIR/sub-mm polarization is due to emission from the aligned dust.
Hence, NIR polarimetry can investigate the magnetic field in regions
where FIR/sub-mm polarimetry is absent,
if background stars exist.
This advantage is clearly shown in Fig. \ref{fig:PABGChuss}.
We have detected polarization at positions
where blue bars are not shown.
The distribution of the position angles including such low emission regions
shows a globally toroidal magnetic configuration at the GC.

\section{SUMMARY}

We have measured the near-infrared polarization of point sources toward the Galactic center (GC)
in the $20\arcmin \times 20 \arcmin$ region centered at Sgr A$^*$.
The difference in the Stokes parameters
between stars at the close and far sides of the GC
reveals the polarization originating from the central 1$-$2 kpc region
of our Galaxy.
The distribution of the position angle for the central region shows
good agreement with those obtained from polarized emission of dust in the GC,
showing that the near-infrared polarization of point sources
can be used as a tool to investigate the magnetic field configuration of the GC.
The position angles have a peak at $\sim 20\degr$,
which is almost parallel to the Galactic plane,
suggesting a global toroidal magnetic field in the region.

\acknowledgements

We are grateful to Hiroshi Akitaya for his helpful comments,
and Jun Hashimoto for his help with our analysis.
We thank the staff of the South African Astronomical Observatory (SAAO)
for their support during our observations.
The IRSF/SIRIUS project was initiated and supported by Nagoya
University and the National Astronomical Observatory of Japan
in collaboration with the SAAO.
SN is financially supported by the Japan Society for the Promotion of Science (JSPS) 
through the JSPS Research Fellowship for Young Scientists.
This work was supported by KAKENHI,
Grant-in-Aid for Young Scientists (B) 19740111,
Grant-in-Aid for Scientific Research (A) 19204018, and
Grant-in-Aid for Scientific Research on Priority Areas 15071204,
and also supported in part 
by Grants-in-Aid for the 21st Century 
COE ``The Origin of the Universe and Matter: 
Physical Elucidation of the Cosmic History'' from the MEXT of Japan.
This publication makes use of data from the Two Micron All Sky Survey, 
a joint project of the University of Massachusetts,
the Infrared Processing and Analysis Center, 
the National Aeronautics and Space Administration, 
and the National Science Foundation.



\begin{thebibliography}{}

\bibitem[Chuss et al.(2003)]{Chuss03}
Chuss, D. T., et al. 2003, \apj, 599, 1116

\bibitem[Davies et al.(1997)]{Davies97}
Davies, J. I., Trewhella, M., Jones, H., Lisk, C., Madden, A., Moss, J.
1997, \mnras, 288, 679

\bibitem[Dotson et al.(2000)]{Dotson00}
Dotson, J. D., Davidson, J., Dowell, C. D., Schleuning, D. A., 
\& Hildebrand, R. H. 2000, \apjs, 128, 335

\bibitem[Eckart et al.(1995)]{Eckart95}
Eckart, A., Genzel, R., Hofmann, R., Sams, B. J., \& Tacconi-Garman, L. E.
1995, \apj, 445, L23

\bibitem[Girardi et al.(2002)]{Girar02}
Girardi, L., Bertelli, G., Bressan, A., Chiosi, C., 
Groenewegen, M. A. T., Marigo, P., Salasnich, B., \& Weiss, A.
2002, \aap, 391, 195

\bibitem[Hildebrand et al.(1993)]{Hildebrand93}
Hildebrand, R. H., Davidson, J. A., Dotson, J., 
Figer, D. F., Novak, G., Platt, S. R., \& Tao, L.
1993, \apj, 417, 565

\bibitem[Goodrich(1986)]{Goodrich86}
Goodrich, R. W. 1986, \apj, 311, 882

\bibitem[Hough et al.(1978)]{Hough78}
Hough, J. H., McCall, A., Adams, D. J., \& Jameson, R. F.
1978, \aap, 69, 431

\bibitem[Kandori et al.(2006)]{Kandori06}
Kandori, R., et al. 2006, Proc. SPIE, 6269, 159

\bibitem[Kobayashi et al.(1980)]{Kob80}
Kobayashi, Y., Kawara, K., Kozasa, T., Sato, S., \& Okuda, J.
1980, \pasj, 32, 291

\bibitem[Kobayashi et al.(1983)]{Kob83}
Kobayashi, Y., Okuda, H., Sato, S., Jugaku, J., \& Dyck, H. M.
1983, \pasj, 35, 101

\bibitem[Lang et al.(1999)]{Lang99}
Lang, C. C., Morris, M., \& Echevarria, L.
1999, \apj, 526, 727

\bibitem[LaRosa et al.(2004)]{LaRosa04}
LaRosa, T. N., Nord, M. E., Joseph, T., Lazio, W., \& Kassim, N. E.
2004, \apj, 607, 302

\bibitem[Launhardt et al.(2002)]{Launhardt02}
Launhardt, R., Zylka, R., \& Mezger, P. G.
2002, \aap, 384, 112

\bibitem[Lebofsky et al.(1982)]{Lebofsky82}
Lebofsky, M. J., Rieke, G. H., Deshpande, M. R., \& Kemp, J. C.
1982, \apj, 263, L672

\bibitem[Maihara et al.(1973)]{Maihara73}
Maihara, T., Okuda, H., \& Sato, S.
1973, in IAU Symposium no. 52,
Interstellar Dust and Related Topics,
ed. Greenberg, J. M., \& van de Hulst, H. C., 191

\bibitem[Mezger et al.(1996)]{Mezger96}	
Mezger, P. G., Duschl, W. J., \& Zylka, R.
1996, \aapr, 7, 289

\bibitem[Morris \& Yusef-Zadeh(1985)]{Morris85}
Morris, M., \& Yusef-Zadeh, F.
1985, \aj, 90, 2511

\bibitem[Morris et al.(1992)]{Morris92}
Morris, M., Davidson, J. A., Werner, M., Dotson, J., Figer, D. F., Hildebrand, R., Novak, G., Platt, S.
1992, \apj, 399, L63

\bibitem[Morris \& Serabyn(1996)]{Morris96}
Morris, M., \& Serabyn, E.
1996, \araa, 34, 645

\bibitem[Morris(1998)]{Morris98}
Morris, M. 1998, in IAU Symp. 184, The Central Regions of the Galaxy and
Galaxies, ed. Y. Sofue (Dordrecht: Kluwer), 331

\bibitem[Nagashima et al.(1999)]{Nagas99} Nagashima, C., et al. 1999,
in Star Formation 1999, ed. T. Nakamoto
(Nobeyama : Nobeyama Radio Obs.), 397

\bibitem[Nagayama et al.(2003)]{Nagay03} Nagayama, T., et al. 2003, 
Proc. SPIE, 4841, 459

\bibitem[Nagata et al.(1994)]{Nagata94}
Nagata, T., Kobayashi, N., \& Sato, S.
1994, \apj, 423, L113

\bibitem[Nishiyama et al.(2006a)]{Nishi06a}
Nishiyama, S., et al. 2006b, \apj, 638, 839

\bibitem[Nishiyama et al.(2006b)]{Nishi06b}
Nishiyama, S., et al. 2006a, \apj, 647, 1093

\bibitem[Novak et al.(2000)]{Novak00}
Novak, G., Dotson, J. L., Dowell, C. D., Hildebrand, R. H., Renbarger, T., \& Schleuning, D. A.
2000, \apj, 529, 241

\bibitem[Novak et al.(2003)]{Novak03}
Novak, G., Chuss, D. T., Renbarger, T., Griffin, G. S., Newcomb, M. G., 
Peterson, J. B., Loewenstein, R. F., Pernic, D., \& Dotson, J. L.
2003, \apj, 583, L83

\bibitem[Ott et al.(1999)]{Ott99} 
Ott, T., Eckart, A., \& Genzel, R. 1999, \apj, 523, 248

\bibitem[Persson et al.(1998)]{Persson98} 
Persson, S. E., Murphy, D. C., Krzeminski, W., Roth, M., \& Rieke, M. J. 
1998, \aj, 116, 2475

\bibitem[Serabyn \& Morris(1996)]{Serabyn96}
Serabyn, E., Morris, M.
1996, \nat, 382, 602

\bibitem[Shibata \& Uchida(1986)]{Shibata86} 
Shibata, K., \& Uchida, Y. 1986, \pasj, 38, 631

\bibitem[Skrutskie et al.(2006)]{Skrutskie06}
Skrutskie, M. F., et al. 2006, \aj, 131, 1163

\bibitem[Sofue \& Handa(1984)]{Sofue84}
Sofue, Y., \& Handa, T. 1984, \nat, 310, 568

\bibitem[Stetson(1987)]{Stetson87}
Stetson, P. B. 1987, \pasp, 99, 191

\bibitem[Tokunaga(2000)]{Tokunaga00}
Tokunaga, A. T. 2000, Astrophysical Quantities, 4th edition, 
ed. Cox, A. (AIP Press), p.143

\bibitem[Tokunaga et al.(2002)]{Tok02}
Tokunaga, A. T., Simons, D. A., \& Vacca, W. D. 
2002,  \pasp, 114,  180

\bibitem[Tsuboi et al.(1986)]{Tsuboi86}
Tsuboi, M., Inoue, M., Handa, T., Tabara, H., Kato, T., Sofue, Y., \& Kaifu, N.
1986, \aj, 92, 818

\bibitem[Uchida et al.(1985)]{Uchida85}
Uchida, Y., Sofue, Y., \& Shibata, K.
1985, \nat, 317, 699

\bibitem[Wainscoat et al.(1992)]{Wainscoat92}
Wainscoat, R. J., Cohen, M., Volk, K., Walker, H. J., \& Schwartz, D.
1992, \apjs, 83, 111

\bibitem[Werner et al.(1988)]{Werner88}
Werner, M. W., Davidson, J. A., Morris, M., Novak, G., Platt, S. R., Hildebrand, R. H.
1988, \apj, 333, 729

\bibitem[Yusef-Zadeh et al.(1984)]{Yusef84}
Yusef-Zadeh, F., Morris, M., \& Chance, D.
1984, \nat, 310, 557

\bibitem[Yusef-Zadeh et al.(2004)]{Yusef04}
Yusef-Zadeh, F., Hewitt, J. W., \& Cotton, W.
2004, \apjs, 155, 421


\end{thebibliography}
\end{document}